\documentclass[a4paper,fleqn,usenatbib]{mnras}

\usepackage{newtxtext,newtxmath}
\usepackage[T1]{fontenc}
\usepackage{ae,aecompl}

\usepackage{graphicx}	% Including figure files
\usepackage{amsmath}	% Advanced maths commands
\usepackage{amssymb}	% Extra maths symbols

\title[Discovery of a 23.8 h QPO in \xmmuj]{Discovery of a 23.8\,h QPO in the SWIFT light curve of \xmmuj}

\author[S. Carpano et al.]{
S. Carpano,$^{1}$\thanks{E-mail:scarpano@mpe.mpg.de}
C. Jin,$^{1},^{2}$
\\
$^{1}$Max-Planck-Institut f\"{u}r extraterrestrische Physik, Giessenbachstra{\ss}e 1, 85748 Garching, Germany\\
$^{2}$National Astronomical Observatories, Chinese Academy of Sciences, A20 Datun Road, Beijing 100101, China
}

\date{Accepted 2018 March 28. Received 2018 March 22; in original form 2018 February 15}

\pubyear{2018}

\newcommand{\xmmuj}{XMMU\,J134736.6+173403}

\begin{document}
\label{firstpage}
\pagerange{\pageref{firstpage}--\pageref{lastpage}}
\maketitle

\begin{abstract}
\xmmuj\ is an X-ray source discovered serendipitously by XMM-Newton which was found to be spatially coincident with a pair of galaxies, including a Seyfert 2 galaxy, but presented in 2003 a very sharp persistent flux drop of a factor 6.5 within 1\,h. From the analysis of a set of 29 Swift observations conducted from the 6 February to the 23 May 2008, we  discovered twin-peak quasi-periodic oscillations (QPOs) with periods of 23.82$\pm$0.07\,h  and 71.44$\pm$0.57\,h. Using a Chandra observation of 2008, we evaluate more accurately the position of the X-ray source  and show that the new source coordinates coincide with the position of the Seyfert 2 galaxy. We provide a detailed spectral energy distribution of the AGN counterpart using multi-wavelength observations. The AGN is radio-loud and the broadband SED modelling indicates a black hole with a mass of $9.8\times10^{6}M_{\odot}$, that accretes at an Eddington ratio of 0.047. QPOs for active galaxies have been reported so far in only few cases, the most reliable one being from RE\,J1034+396 for which a 1\,h periodicity has been discovered analysing a $~$91\,ks XMM-Newton observation. Twin peak QPOs with an observed frequency ratio of 3:1 have not been reported so far for any AGN. From resonance models of the epicyclic frequencies we evaluate the different possible mass-spin relations. It's still not clear what could have been the origin of the high flux and sharp drop only observed in 2003.
\end{abstract}


%%%%%%%%%%%%%%%%%%%%%%%%%%%%%%%%%%%%%%%%%%%%%%%%%%

%%%%%%%%%%%%%%%%% BODY OF PAPER %%%%%%%%%%%%%%%%%%

\section{Introduction}
Quasi-periodic oscillations (QPOs) are found in all kind of accreting systems that include white dwarfs, neutron stars or black holes \citep[e.g.][]{Warner2003, vanderklis2006}. It is commonly accepted that QPOs come from the inner region of the accretion disk, in the vicinity of the compact object. Active galaxies are believed to be scaled-up version of galactic black holes (GBH). The presence of quasi-periodic oscillations (QPOs) in Seyfert galaxies was already addressed by \cite{Vaughan2005}, since they possess the same noise power spectra as galactic black holes but with lower frequencies, scaling inversely with the mass of the black hole \citep{Remillard2006}. QPOs in GBHs are divided into 2 groups, low-frequency QPOs (LF QPOs), with frequencies from few mHz to 30 Hz and the high frequency QPOs (HF QPOs) with frequencies above 100 Hz.
LF QPOs with $f\sim$\,1Hz in a GBH with a typical mass of $\sim$10M$_{\odot}$, would be translated for a Seyfert galaxy to $f_\textrm{LFQPO}\sim$10$^{-5} (M_\textrm{BH}$/10$^6$ M$_{\odot})^{-1}$\,Hz, i.e at time scales >100\,ks \citep{Vaughan2005} and are therefore difficult to be detected with observatories like XMM-Newton or Chandra which have short exposures. For HF QPOs time scales are shorter, above 400 s \citep{Vaughan2005}.
 
A few years later, \cite{Gierlinski2008} reported for the first time the discovery of a robust 1hr QPO of a Seyfert 1 galaxy, RE J1034+396, located at z=0.042, using 91 ks XMM-Newton observation. Its black hole mass has not yet been constrained and, depending on the method used, it should lie between 10$^6$ and 10$^7$ M$_{\odot}$ \citep{Czerny2016}.  Another, more recent, possible detection of a QPO in a AGN was reported by \cite{Pan2016}, for another Seyfert 1 galaxy, 1H 0707-495 located at z=0.04, with a mass estimated by the authors of 5.2$\times$10$^6$ M$_{\odot}$, also using a long 100 ks XMM-Newton observation. For both systems the QPOs were not found in all available XMM-Newton observations but only in some of them suggesting that these periodic modulations are transient, and may have a dependence on the spectral state \citep{Alston2014}.
 
\xmmuj\ is an X-ray source discovered serendipitously by XMM-Newton which is spatially coincident with a pair of galaxies \citep{Carpano2008}, both located at redshift z=0.045. The authors reported in an XMM-Newton observation from 2003 a very sharp flux drop by a factor 6.5 within 1\,h,  followed by a persistent low-flux state. This is difficult to be attributed to any AGN activity. In this paper we analyse Chandra and Swift observations that were performed following the publication of the source discovery and the more recent XMM-Newton observations. 

The article is organised as follows. In paragraph \ref{sec2}, we describe the Chandra, Swift and XMM-Newton observations and detail the analysis of the X-ray data and QPO search. We then build the spectral energy distribution (SED) and show the results in paragraph \ref{sec-sed}. In the next section we discuss the black hole mass and  the mass-spin relation. Conclusions are ending the paper.

%__________________________________________________________________

\section{Analysis of the Chandra, Swift and XMM-Newton observations}
\label{sec2}

\subsection{Accurate source position using Chandra data}
\begin{figure}
   \centering
    \resizebox{\hsize}{!}{ \includegraphics{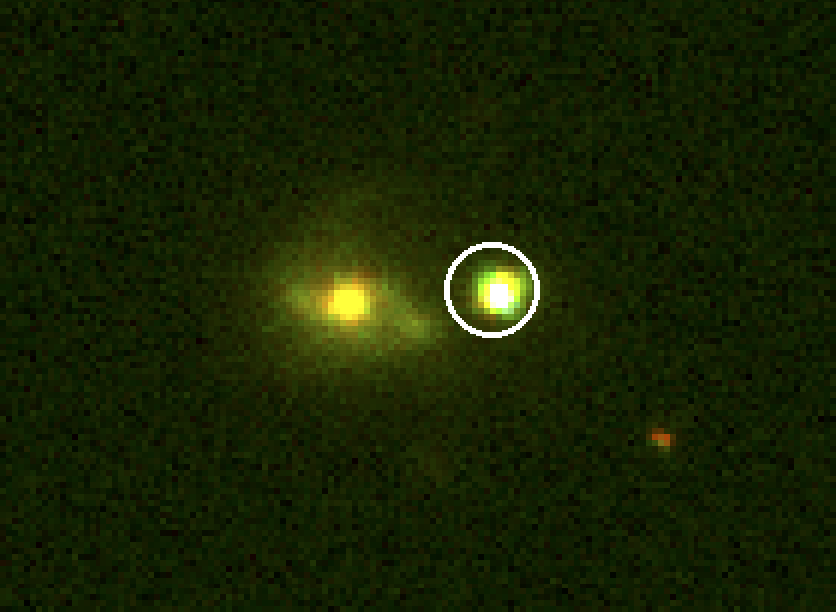}}
   \caption{RGB image from the Sloan Digital Sky Survey (SDSS), where red colour is associated with the R band, green colour with the G band and blue colour with the U band. The circle is centred on the position of the X-ray source determined by Chandra data with a radius of 3$''$.}
              \label{fig_optic}
 \end{figure}

We use the unique Chandra observation targeted on \xmmuj\ (i.e. with the source on-axis), performed on the 29th November 2008, to evaluate the position of the X-ray source more accurately. The data were recorded with the ACIS-S instrument, in imaging (FAINT) mode, for a short 2\,ks exposure. The \texttt{\textbf{ciao}} tool \texttt{celldetect}, provides the new coordinates $\alpha_\text{J2000}=13^\text{h}47^\text{m} 36\fs{}43$ and $\delta_\text{J2000}=+17^\circ 34' 04\farcs 84$ with a statistical error of $0\farcs 05$, which is negligible with respect to the absolute astrometric accuracy of $\sim 0\farcs 6-0\farcs 8$ . It is now clear from Fig.~\ref{fig_optic} that the X-ray source is associated with the Seyfert 2 galaxy, also known as SDSS J134736.39+173404.6, and can't be attributed to a foreground X-ray binary as suggested by \cite{Carpano2008}.

\subsection{Swift observations and QPOs search}
\label{sec:swift}

A set of 29 Swift observations, pointing on \xmmuj\ and listed in Table\,\ref{tab:swift}, were performed between the 6 February to the 23 May 2008 with XRT exposures varying from $~$800 to $~$8000\,s. Source and background extraction regions are centred around the coordinates obtained from the Chandra data. The source region is a circle with a radius of 40$''$ and the background region an annulus with inner and outer radius of 50$''$ to 70$''$, respectively. For every observation, we extract the number of source and background counts obtained in every good time interval (GTI) defined in the file header. Net count rates are then obtained by subtracting the background counts to the source counts after scaling for the respective areas and by dividing by the GTI length.

Columns 1 to 4 of Table\,\ref{tab:swift} show the observation ID, the start of the observation (date and MJD), and the exposure time (in seconds), taken from the \texttt{ONTIME} keyword. 

\begin{table}
 \centering
 \caption{Consecutive observations targeted on \xmmuj\  performed with Swift used for the QPO analysis (see text for more details).} 
  \label{tab:swift}

\begin{tabular}[t]{| c c c c |}
\hline
obs ID & Date  & MJD & Ontime\\
  &   &  & (s) \\
\hline
00031104001 & 2008-02-06T15:01:20 & 54502.63 & 2096 \\ 
00031104002 & 2008-02-07T10:13:57 & 54503.43 & 787 \\ 
00031104003 & 2008-02-09T14:59:00 & 54505.63 & 1978 \\ 
00031104004 & 2008-02-10T02:15:06 & 54506.09 & 7196 \\ 
00031104005 & 2008-02-12T05:38:22 & 54508.24 & 3824 \\ 
00031104006 & 2008-02-13T00:57:18 & 54509.04 & 968 \\ 
00031104007 & 2008-02-14T01:09:30 & 54510.05 & 3282 \\ 
00031104008 & 2008-02-15T04:52:47 & 54511.20 & 3475 \\ 
00031104009 & 2008-02-16T01:34:26 & 54512.07 & 4776 \\ 
00031104010 & 2008-02-17T03:22:54 & 54513.14 & 3718 \\ 
00031104011 & 2008-02-18T03:28:34 & 54514.15 & 3668 \\ 
00031104012 & 2008-03-02T10:46:19 & 54527.45 & 4054 \\ 
00031104013 & 2008-03-03T09:31:42 & 54528.40 & 1241 \\ 
00031104014 & 2008-03-04T01:31:37 & 54529.06 & 2906 \\ 
00031104015 & 2008-03-05T14:29:35 & 54530.60 & 3831 \\ 
00031104016 & 2008-03-06T17:56:20 & 54531.75 & 2247 \\ 
00031104017 & 2008-03-15T14:09:22 & 54540.59 & 1138 \\ 
00031104018 & 2008-03-17T11:05:23 & 54542.46 & 1700 \\ 
00031104019 & 2008-03-23T09:38:34 & 54548.40 & 3192 \\ 
00031104020 & 2008-03-29T02:07:02 & 54554.09 & 2605 \\ 
00031104021 & 2008-04-04T00:00:57 & 54560.00 & 2615 \\ 
00031104022 & 2008-04-18T06:19:07 & 54574.26 & 2084 \\ 
00031104023 & 2008-04-22T12:33:43 & 54578.52 & 2971 \\ 
00031104024 & 2008-04-25T00:16:01 & 54581.01 & 2959 \\ 
00031104025 & 2008-04-28T01:53:23 & 54584.08 & 3676 \\ 
00031104028 & 2008-05-07T10:52:29 & 54593.45 & 2615 \\ 
00031104029 & 2008-05-10T01:30:29 & 54596.06 & 2868 \\ 
00031104030 & 2008-05-13T00:16:29 & 54599.01 & 3122 \\ 
00031104031 & 2008-05-23T14:03:40 & 54609.59 & 7998 \\ 
\hline

\end{tabular}
\end{table}

\begin{figure}
\centering
\resizebox{\hsize}{!}{\includegraphics{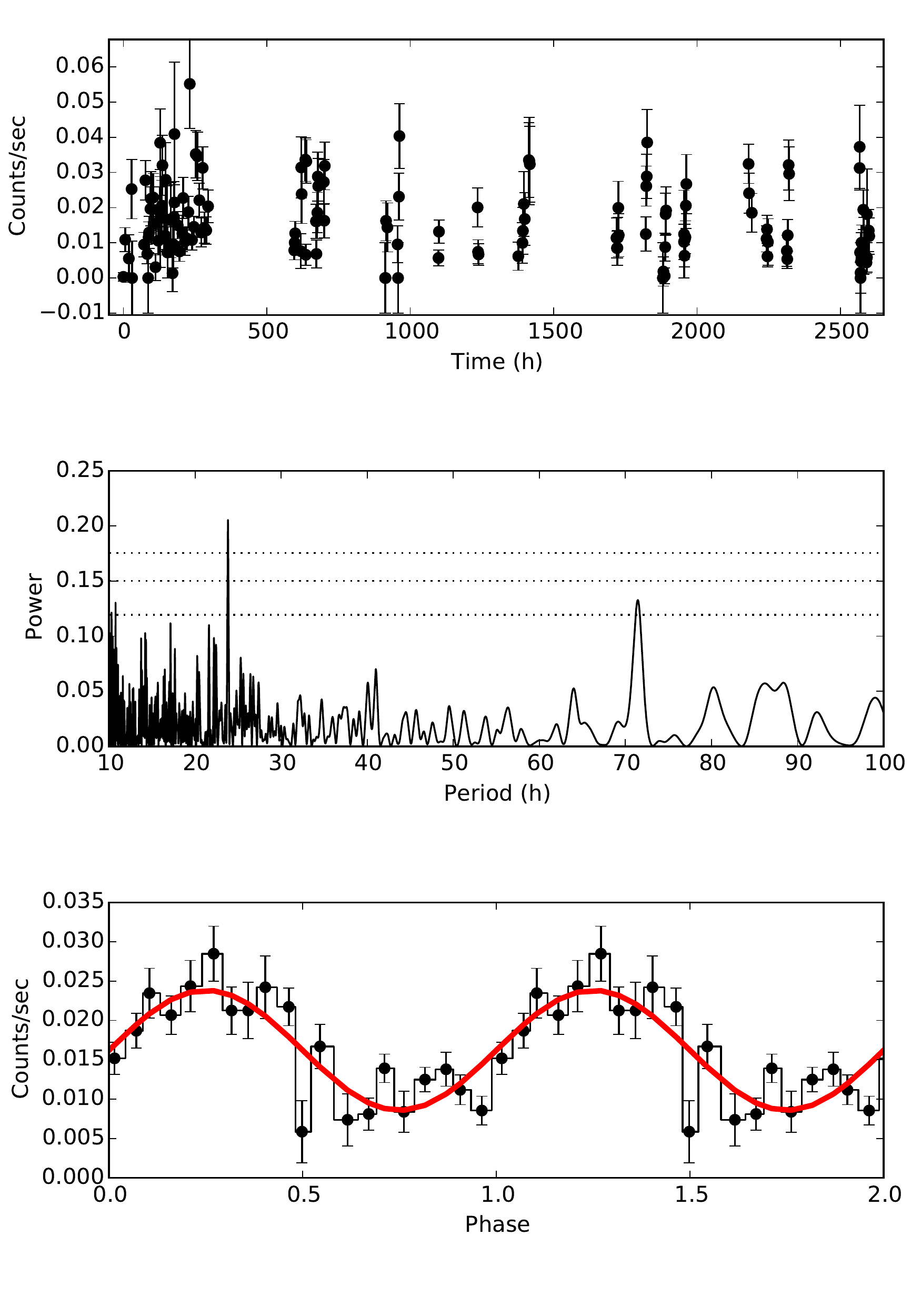}}
  \caption{Top: background subtracted XRT Swift light curve of \xmmuj\  in full  (0.2--10\,keV) energy band. Middle: Lomb Scargle periodogram with white-noise confidence levels at 90$\%$, 99$\%$ and 99.9$\%$, where a clear peak appears at 23.82\,h and a second one at three times the period. Bottom: folded light curve with fitted sine curve over plot.}
        \label{fig_swift}
\end{figure}

We then look for periodic signals via Lomb-Scargle periodogram analysis \citep{Lomb1976, Scargle1982}, in the full (0.2--10\,keV) energy band and in the period range from 10 to 100\,h, using the \texttt{Python} algorithm from the \texttt{astroML} library. The results of the analysis is shown in Fig.~\ref{fig_swift} were the original light curve is shown at the top, the periodogram in the middle and the folded light curve at the bottom. A clear peak is visible at a period 23.82\,h with an uncertainty of 0.07\,h which corresponds to 1$\sigma$ of the gaussian function fitted around the peak.  We can also see a second peak at exactly three times the period (P=71.44$\pm$0.57\,h), but with a lower amplitude. 
The ratio between the two peak heights ($h_{24}/h_{71}$) depends on the energy: while it is of 1.5 in the full (0.2--10\,keV) band, it increases to 1.8 in the soft (0.2--1.5\,keV) band and decreases to 0.8 in the hard (1.5--10\,keV) band.

We calculated confidence levels by bootstrapping the data 10$,$000 times and evaluate the maximum of the corresponding periodograms. 
Since the power spectrum is flat, we assume white noise is dominating the data.
The signal at 23.82\,h has a confidence level above 99.9$\%$, while the second peak is only above 90$\%$. 
The bottom part of the figure shows the folded light curve and a sine function over plotted, with phase 0 corresponding to MJD=54502.9402746 (maximum of the function at phase 0.75). The amplitude of the sine function is 0.008 counts/s and the average count rate is 0.016 counts/s, which is well below the Swift XRT count rate for pile-up in photon counting mode (0.6 counts/s\footnote{\url{http://www.swift.ac.uk/analysis/xrt/xrtpileup.php}}).

%We also checked that the peaks were not produced by any windowing effect by evaluating the Lomb Scargle periodogram for the background only, which is extracted over a region that is large enough to have a similar count rate as of the source. None of the 23.8\,h and 71.4\,h peaks appear in the periodogram of the background light curve.

\subsection{Spectral analysis of the XMM-Newton observations}
There are currently 9 XMM-Newton observations of \xmmuj, the first, with ID 0144570101 and published by \cite{Carpano2008}, was performed on the 24 June 2003 with a total exposure time of 66\,ks. A detail of all XMM-Newton observations, where the source is in the field-of-view, are given in Table~\ref{tab:xmmobs}.  A set of shorter exposures, from 10 to 15\,ks, were recorded between June 2010 and January 2012. The data are reduced following standard procedures using the XMM-Newton SAS data software version 16.1. No background flares were present during these observations.  Single to quadruple events are used for MOS \citep{Turner2001} cameras, with FLAG=0, while, except for the first observation, the source is out of the field of view of the pn camera \citep{Strueder2001} (exposures taken in large window mode).

Spectra are modelled using the $XSPEC$ (v12.9.1m) software, and as for the first observation \citep{Carpano2008}, a simple absorbed power-law model is sufficient to model the data (\texttt{phabs*power}). Since the value for the $N_\textrm{H}$ could not be constrained properly, it is frozen to the Galactic hydrogren column density measured in that direction of the sky. We consider here the sum of both atomic $N_\textrm{H$_1$}$ and molecular  $N_\textrm{H$_2$}$ components as suggested by \cite{Willingale2013}, which is of $N_\textrm{H$_\textrm{tot}$}$=1.9$\times 10^{20}\,\text{cm}^{-2}$ using the web interface (\url{http://www.swift.ac.uk/analysis/nhtot/index.php}).
%%, i.e. 1.8$\times 10^{20}\,\text{cm}^{-2}$ \citep{Dickey1990, Kalberla2005} using the web interface (\url{https://heasarc.gsfc.nasa.gov/cgi-bin/Tools/w3nh/w3nh.pl}). 
Table~\ref{tab:spec_fit} shows the results of the spectral fit, giving the value of the power-law index, the absorbed flux in the 0.3-10\,\textrm{keV} band,  the reduced chi-square and the number of degrees of freedom.
When the spectrum of more than one camera is available, the two or three spectra are fitted simultaneously without the addition of a floating cross-normalisation component.
 
\begin{table}
 \centering
 \caption{Summary of the XMM-Newton observations  and exposure ID where \xmmuj\ is in the field-of-view.} 
  \label{tab:xmmobs}

\begin{tabular}[t]{| l l l l l |}
\hline
Obs ID & Date start & ExpID  & Ontime & Filter\\ 
\hline 
0144570101 & 2003-06-24 & PNS003 &  62584  & Thick \\ 
& & M1S001 & 60323 & Thick \\ 
& & M2S002 & 60320 & Thick \\ 
0651140201 & 2010-06-19 & M1S001 &  14588  & Thick \\ 
& & M2S002 & 14600 & Thick \\ 
0651140301 & 2010-07-23 & M1S001 &  9588  & Thick \\ 
& & M2S002 & 9577 & Thick \\ 
0651140401 & 2010-12-19 & M2S002 &  9577  & Thick \\ 
0651140501 & 2011-01-22 & M2S002 &  13190  & Thick \\ 
0671150501 & 2011-06-19 & M1S001 &  11592  & Thick \\ 
& & M2S002 & 11591 & Thick \\ 
0671150601 & 2011-07-15 & M1S001 &  12400  & Thick \\ 
& & M2S002 & 12400 & Thick \\ 
0671150701 & 2011-12-24 & M2S002 &  11000  & Thick \\ 
0671150801 & 2012-01-20 & M2S002 &  10600  & Thick \\ 
\hline

\end{tabular}
\end{table}

\begin{table}
 \centering
 \caption{Results of the spectral fits of \xmmuj\  from all existing XMM-Newton observations, using an absorbed power-law model (\texttt{phabs*power}, in XSPEC, with fixed value for N$_{\rm H}$=1.8$\times 10^{20}\,\text{cm}^{-2})$. The high and low-state flux during observation  0144570101 is 13.3 and 2.4$\times 10^{-13}$ $\,\text{ergs}\,\text{s}^{-1}\,\text{cm}^{-2}$, respectively \citep{Carpano2008}} 
  \label{tab:spec_fit}

\begin{tabular}[t]{| l l l l l |}
\hline
obs ID & $\Gamma$ & $F_{0.3-10\,\textrm{keV}}$ & $\chi_{\textrm{red}}$ & dof\\
  &   & $\times 10^{-13}$ (\,cgs) & & \\
\hline
0144570101& 2.69$^{+0.03}_{-0.03}$ & 6.06$^{+0.06}_{-0.08}$ & 1.19 & 452\\ 
0651140201& 2.80$^{+0.23}_{-0.22}$ & 2.35$^{+0.20}_{-0.15}$ & 1.08 & 24\\ 
0651140301& 2.48$^{+0.15}_{-0.15}$ & 5.61$^{+0.25}_{-0.36}$ & 1.15 & 38\\ 
0651140401& 2.58$^{+0.25}_{-0.24}$ & 3.87$^{+0.32}_{-0.29}$ & 1.17 & 17\\ 
0651140501& 2.78$^{+0.24}_{-0.24}$ & 2.74$^{+0.22}_{-0.20}$ & 1.00 & 18\\ 
0671150501& 2.79$^{+0.16}_{-0.16}$ & 4.67$^{+0.26}_{-0.17}$ & 1.07 & 41\\ 
0671150601& 2.59$^{+0.15}_{-0.15}$ & 5.22$^{+0.31}_{-0.18}$ & 1.28 & 42\\ 
0671150701& 2.55$^{+0.17}_{-0.17}$ & 4.99$^{+0.37}_{-0.28}$ & 0.67 & 25\\ 
0671150801& 2.55$^{+0.15}_{-0.14}$ & 6.47$^{+0.38}_{-0.29}$ & 0.88 & 36\\

\hline

\end{tabular}
\end{table}

As for observation 0144570101 from 2003, the spectra are quite soft and are not varying significantly from one exposure to another. The flux from the observations performed between June 2010 to January 2012 is varying by maximum a factor of 3 but is never reaching the high flux of 1.33$\times 10^{-12}$ $\,\text{ergs}\,\text{s}^{-1}\,\text{cm}^{-2}$  measured at the first half of the first observation \citep{Carpano2008}.

\section{Spectral Energy Distribution}
\label{sec-sed}
\begin{figure}
%\resizebox{\hsize}{!}{\includegraphics[bb=50 210 558 690]{bestfit1_plot.ps}}
\resizebox{\hsize}{!}{\includegraphics{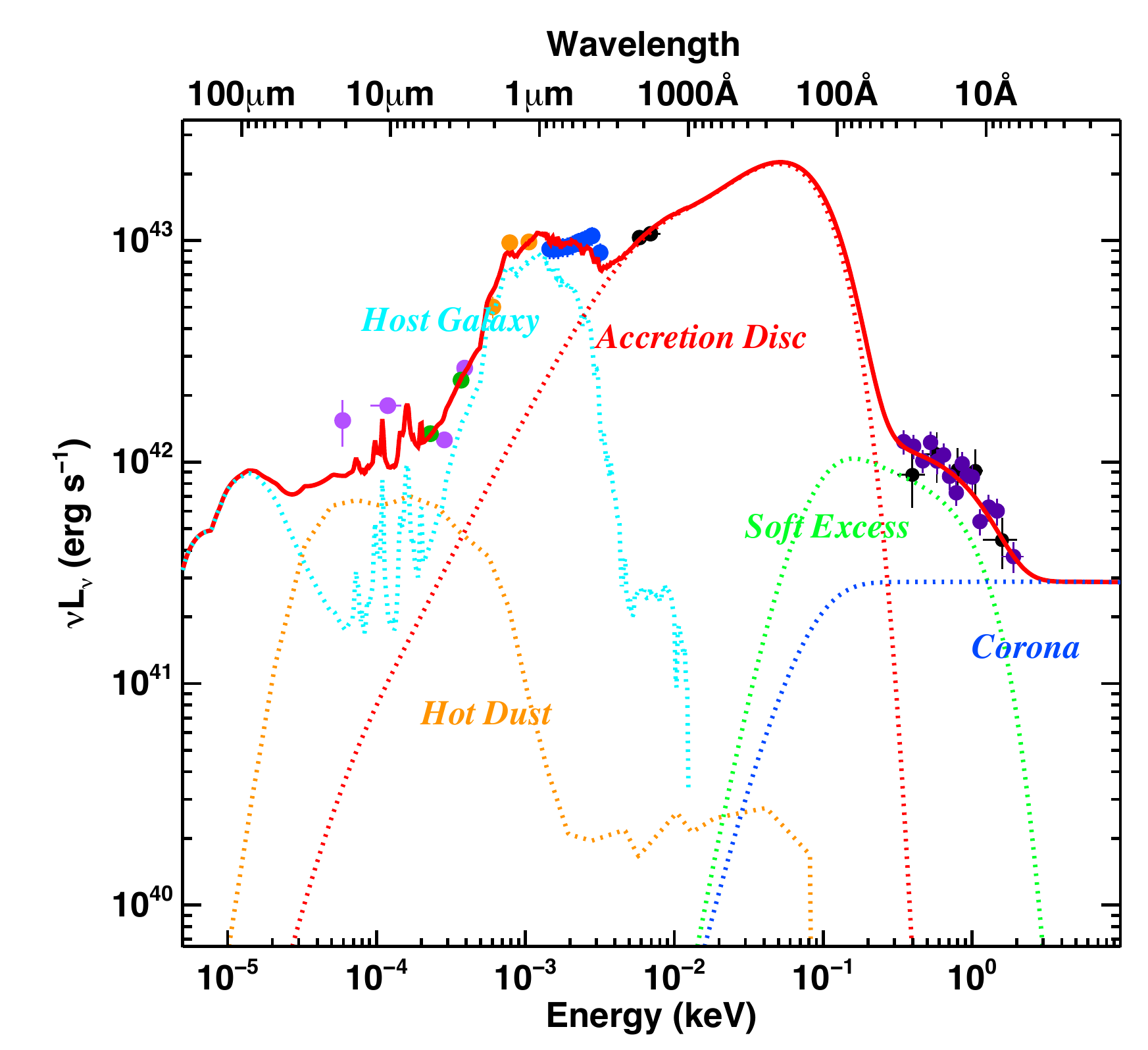}}
\caption{The unabsorbed broadband spectral energy distribution of \xmmuj. The data sets include Swift XRT (black points in the X-ray), UVOT (black points in the UV), XMM-Newton EPIC-pn (purple points, scaled up by 2.03), SDSS (blue points, scaled up by 1.27), 2MASS (orange points, scaled up by 1.27), Spitzer IRAC (green points in the infrared, scaled up by 1.04), WISE (magenta points in the infrared, scaled down by 0.61). The scaling factor is to account for the aperture effect and long-term variability. The SED model is {\sc optxagnf} plus host galaxy and hot dust templates, with different components labelled in the figure.}
\label{fig-sed}
\end{figure}
In order to obtain a more global understanding of the source, we used multi-wavelength archival data to rebuild its broadband SED \citep[see e.g.][]{Jin2016, Jin2017b}. Since the source underwent a sharp flux drop in the X-ray band in June 2003 (during the XMM-Newton observation 0144570101), it is very important to have simultaneous data for optical/UV and X-rays. Unfortunately, there was no OM data for any of the XMM-Newton observations due to the bright star Tau Bo\"{o}tis in the OM field-of-view. For the 29 Swift observations, only 6 have simultaneous exposures in UVOT, and only the observation performed on the 23 May 2008 (with obsID 00031104031) has simultaneous exposures in at least two filters of UVOT (UVM2 and UVW2). For the other observations, the filter was in the "BLOCKED" position, likely due to the presence of Tau Bo\"{o}tis, as for the  XMM-Newton OM telescope. Therefore, we chose observation 00031104031 as the basis for building the broadband SED. Since the signal-to-noise of the XRT spectrum is low in that observation, we also extracted the XMM-Newton EPIC-pn spectrum for the low-state period in June 2003, in order to have more constraints in the X-rays. In the optical band, we used the underlying continuum from the SDSS spectrum taken on the 13 May 2007.

In the infrared band, we included 3 photometric points from 2MASS (J, H, K band, observed on 1998-05-02), 2 photometric points from Spitzer IRAC at 3.6 and 5.8 $\mu$m (observed on 2008-07-22) and 4 photometric points from WISE (band 1-4, averaging over 17 exposures between 2010-01-07 and 2010-07-05). In the radio band, the source was detected by the NRAO/VLA Sky Survey (NVSS, conducted between 1993-09 and 1996-10, Condon et al. 1998) with a flux of 13.0$\pm$0.6 mJy at 1.4 GHz. Then this gives a radio-loudness of $R=38.9$, making it a radio-loud source according to the AGN radio-loudness classification \citep{Kellermann1989}.

All the datasets from infrared to X-ray were converted to PHA files to be imported and modelled in {\sc xspec}.

\begin{table}
\centering
\caption{Best-fit parameters for the {\sc optxagnf} SED model in Fig.~\ref{fig-sed}. Error bars indicate the 90\% confidence ranges. N$_{\rm H, gal}$ and N$_{\rm H, agn}$ are the hydrogen column density from our Milky Way and the AGN host galaxy. r$_{\rm cor}$ is the corona radius. $\tau$ and kT$_{\rm e}$ are the optical depth and electron temperature of the soft X-ray Comptonisation component. {\it fpl} is the energy fraction of the hard X-ray Comptonisation component in the total soft and hard X-ray coronal emission.}
\label{tab-sedfit}
\begin{tabular}{lcc}
\hline
Parameter & Unit & Value \\
\hline
$N_{\rm H, gal}$ & ($10^{20}$ cm$^2$) & 1.90 (fixed) \\
$N_{\rm H, agn}$ &($10^{20}$ cm$^2$) & 0.00$^{+0.02}_{-0.00}$ \\
BH Mass & (10$^6$ M$_\odot$)  & 9.8$^{+18.4}_{-3.6}$\\
L/L$_{\rm Edd}$ & & 0.047$^{+0.062}_{-0.039}$\\
r$_{\rm cor}$ & (r$_{\rm G}$)  & 9.9$^{+7.2}_{-1.2}$ \\
kT$_{\rm e}$ &  (keV) & 0.26$^{+1.33}_{-0.12}$\\
$\tau$ & & 17.3$^{+82.7}_{-12.1}$\\
$\Gamma$ & &  2.0 (fixed) \\
{\it fpl} & & 0.46$^{+0.33}_{-0.46}$\\
\hline
\end{tabular}
\end{table}

We followed the method of \cite{Jin2017b} to model the broadband SED with several components. We used the energy-conserved {\sc xspec } {\tt optxagnf} model to fit the AGN emission from optical to hard X-rays \citep{Done2012, Jin2012a, Jin2012c, Jin2013, Jin2016}. Then we adopted a Sa-type galaxy template \citep[SWIRE library][]{Polletta2007} to model the host galaxy emission from infrared to near-UV%\footnote{There is a normal galaxy, namely SDSS J134737.10+173404.0, located 10 arcsec away from this source at the same redshift of 0.045 (or separated by 9 kpc at the source distance of 184.4 Mpc). This galaxy appears extended in the SDSS image, so it might contribute a small fraction of star-light in the optical spectrum of this source, but this little contamination should not affect our study.}.
\footnote{The companion galaxy, namely SDSS J134737.10+173404.0, visible in Fig.~\ref{fig_optic},  located 10 arcsec away from the AGN and at the same redshift of 0.045 (or separated by 9 kpc at the source distance of 184.4 Mpc),  might contribute for a small fraction of the star-light in the optical spectrum, but this little contamination should not affect our study.}.
 A hot dust template \citep{Silva2004} was also adopted to account for the infrared emission from the AGN torus. Then two sets of extinctions were included in the model. One set is for the extinction in the Milky Way, with $N_{\rm H}=1.9\times10^{20}$cm$^{-2}$ including both gas and molecular hydrogen column densities (Willingale et al. 2013), and $E(B-V)=1.7\times10^{-22}N_{\rm H}$ \citep{Bessell1991}. These are modelled with the {\sc xspec} models {\tt wabs} and {\tt redden}. The other set is for the host galaxy extinction, for which we used the {\tt zwabs} and {\tt zredden} models and the same gas-to-dust relation, but allowed the $N_{\rm H}$ value to be a free parameter. Another issue is the normalisation discrepancy between different datasets, which is mainly due to different aperture sizes of different instruments and the non-simultaneous observations. We adopted a constant as a free parameter to count for these normalisation differences.

In {\sc optxagnf}, we allowed the black hole mass and mass accretion rate to be free parameters. The source co-moving distance was found to be 184.4 Mpc for a flat universe model \citep{Wright2006}. Since there are no valid data points above 2 keV for that spectrum, due to the low count rate, we fixed the photon index at 2.0 \citep{Jin2012c, Martocchia2017}. The best-fit SED is presented in Fig.~\ref{fig-sed} together with the best-fit parameters listed in Table~\ref{tab-sedfit} . It is clear that the soft X-ray emission is dominated by a soft X-ray excess component, the far-UV emission is dominated by the accretion disc, the optical and near-UV emission is mainly from the host galaxy, and the infrared is from both host galaxy and hot dust in the torus. This is consistent with the source being a Seyfert 2 galaxy. The best-fit model has a black hole mass of $9.8\times10^{6}M_{\odot}$, and an Eddington ratio of 0.047, so it is a low mass accretion rate AGN. The soft X-ray Comptonisation component has an electron temperature of 0.26 keV and optical depth of 17.3. Some small scaling factors are required to adjust the normalisation for different datasets, as shown in Fig.~\ref{fig-sed}. The best-fit $\chi^2=263$ for 86 degrees of freedom is relatively large, which is mainly due to the small error bars in the infrared to UV data. This is probably due to the fact that the templates of host galaxy and hot dust are not good enough for this source, or because the long-term variability is more complex than a normalisation change (e.g. the spectral shape also varies). 

Although no host galaxy extinction is required by the model, it does not mean that the X-ray emission must be unobscured. The absorption might be too complicated to be modelled by the neutral absorption model ({\sc wabs}). {The best-fit model provides a corona radius of 9.9 $r_{\rm G}$ (where $r_{\rm G}$ is the Gravitational radius)}, which is much smaller than the mean value of 60 $r_{\rm G}$ found by \cite{Jin2012a} for typical BLS1s, indicating that maybe the X-ray emission in the low-state of observation 0144570101 is not intrinsically weak. Indeed, if the sharp drop is due to an X-ray obscuration event caught in the act, then the intrinsic X-ray luminosity can be a factor of 6.5 higher, the corona radius would be more similar to that of a typical BLS1 \citep{Jin2012c}.

\section{Discussion}
\subsection{Black Hole Mass}
\label{sec-mass}
We have found a black hole mass of $9.8\times10^{6}M_{\odot}$ for \xmmuj\  from the broadband SED fitting, but this value is model dependent, and its systematic uncertainties are difficult to estimate, mainly due to the lack of direct observational constraints on the intrinsic disc emission in the optical/UV band. The popular method of using the single-epoch optical broad-line width to estimate the black hole mass is not valid for this source either, because it is a Seyfert 2 galaxy (i.e. no broad-line component) and its optical emission is dominated by the star-light in the host galaxy.

Another method is to use the scaling relation between the hard X-ray variability and the black hole mass \citep{Miniutti2009, Ponti2012, Jin2016}. However, there were not enough counts in the hard X-rays above 2 keV, for all observations, to allow a useful spectrum or light curve analysis. Therefore, we have to use the soft X-ray band variability as an approximation for the hard X-ray variability, which is certainly not accurate \citep[e.g.][]{Jin2017a}. For all the XMM-Newton observations, we extracted the 0.3-2 keV light curves and calculated the intrinsic excess variability \citep[$\sigma_{rms}^{2}$, ][]{Edelson2002, Vaughan2003, Uttley2014}. For the low-state XMM-Newton observation showing the strongest source variability (ObsID: 0651140401), we found $\sigma_{rms}^{2}=0.119\pm0.086$. The source also appears variable during the high-flux state, so we extracted the first 20 ks of ObsID 0144570101, and found $\sigma_{rms}^{2}=0.019\pm0.006$. For the entire 60 ks of this observation including the sharp flux drop, we found $\sigma_{rms}^{2}=0.031\pm0.006$. These variabilities all correspond to a black hole mass of the order of $10^{6}M_{\odot}$ according to the scaling relations reported by \cite{Ponti2012}. However, this scaling relation has a relatively large intrinsic scatter. Besides, all the other observations have no detectable significant intrinsic variability (partly due to the low source count rate). So we conclude that the X-ray variability can only provide a lower limit for the black hole mass of this source, and it can support any black hole mass above ${10^6}M_{\odot}$.

\subsection{Mass-spin relation}
\label{sec-spin}
Double peak QPOs have been observed for several microquasars like GRO1655--40, XTE 1550--564, H 1743--322, GRS 1915+105 but also for Sgr A$^\star$ \citep[see][and references therein]{Abramowicz2004, Torok2005, Torok2005b, Bambi2017}. Their upper $\nu_U$ and lower  $\nu_L$ frequencies are found in a 3:2 ratio suggesting the presence of resonance of epicyclic frequencies  in the accretion disk. More explicitly, the equatorial circular orbits of a test particle are characterized by the orbital frequency (or Keplerian frequency) $\nu_\phi$ (or $\nu_\textrm{K}$), which is the inverse of the orbital period, the radial epicyclic frequency $\nu_r$, which is the frequency of radial oscillations around the mean orbit and the vertical epicyclic frequency $\nu_\theta$ , which is the frequency of vertical oscillations around the mean orbit \citep{Bambi2017}.

\begin{figure}
\centering
\resizebox{\hsize}{!}{\includegraphics{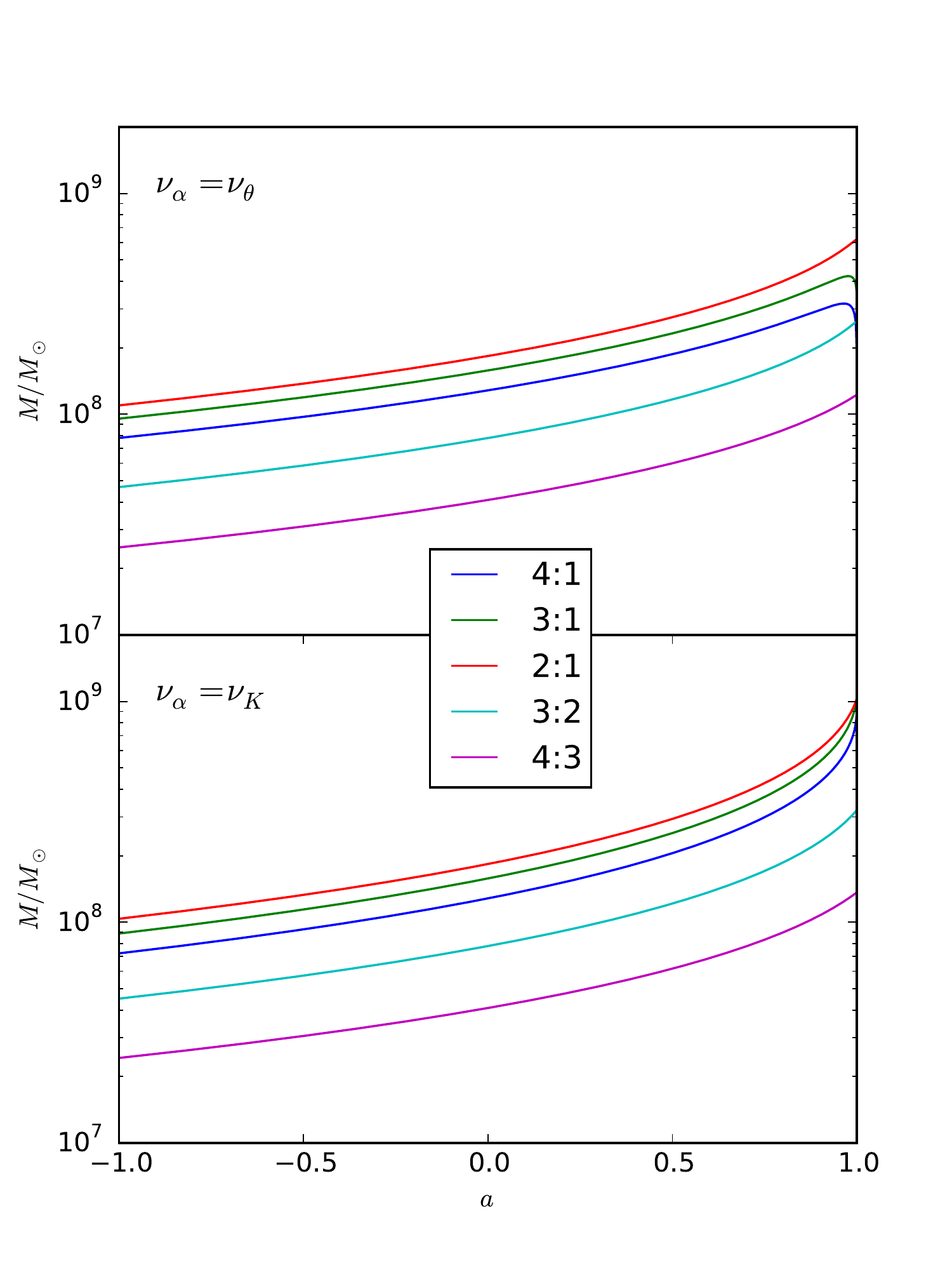}}
  \caption{Mass-spin dependence for the direct and simple combinational resonances matching the observed 3:1 frequency ratio, for  $\nu_{\alpha}=\nu_{\theta}$ (top) and $\nu_{\alpha}=\nu_{K}$ (bottom).}
        \label{fig:mass_spin}
\end{figure}

For Kerr black holes,  the formulae are \citep[see e.g. ][]{Torok2005}:

\begin{align}
\nu_{K}  &=  {1\over 2\pi}\left ({{GM}\over {r_G^{~3}}}\right )^{1/2}\left( x^{3/2} + a \right)^{-1}={1\over 2\pi}\left ({{c^3}\over {GM}}\right )\left( x^{3/2} + a \right)^{-1}, \\
\nu_{\rm r}^2  &=  {\nu_{K}^2}\,\left(1-6\,x^{-1}+ 8 \,a \, x^{-3/2} -3 \, a^2 \, x^{-2} \right), \\
\nu_{\theta}^2  &=  {\nu_{K}^2}\,\left(1-4\,a\,x^{-3/2}+3a^2\,x^{-2}\right),
\end{align}

\noindent where G is the gravitational constant, c the speed of light,  $x = r/r_\textrm{G}$ is the dimensionless radius expressed in terms of the gravitational radius, r$_\textrm{G}=  (GM/c^2)$, of the black hole, and a = Jc/G$M^2$ is the dimensionless spin.

To explain the 3:2 ratio between the two observed frequencies, resonance models have been used. For parametric resonances, where ${\nu_r \over \nu_{\theta}}={2\over n}$ (n=1, 2, 3,..), with the condition that $\nu_{\theta}$>$\nu_r$, the resonances are naturally explained by associating $\nu_U$ with  $\nu_{\theta}$ and $\nu_L$ with  $\nu_r$. For the forced resonances, the 3:2 ratio are explained by associating $\nu_U$ with  $\nu_{\theta}$ and $\nu_L$ with  $\nu_{\theta}-\nu_r={2\over3} \nu_{\theta}$ (for the 3:1 epicyclic frequency resonance) or by associating $\nu_U$ with  $\nu_{\theta}+\nu_r={3\over2} \nu_{\theta}$ and $\nu_L$ with $\nu_{\theta}$ (for the 2:1 epicyclic frequency resonance).

\cite{Zhang2017} recently claimed the discovery of a twin frequency for the Seyfert 1 galaxy Mrk 766, in the ratio 3:2, although both QPOs were not observed simultaneously.
For  \xmmuj\  the second QPO period is at 71.44$\pm$0.57\,h which makes the ratio between the two observed frequencies exactly 3. The 3:1 ratio for the observed frequencies is more rare in the literature. \cite{Stuchlik2013} proposed multi-resonance models for QPOs that explain any ratio between the observed frequencies by performing several possible combinations of the epicyclic frequencies.

In this paper, we investigate only direct resonances and simple combinational resonances summarized in
Table~\ref{tab:reson}, matching the observed 3:1 frequency ratio (i.e. $\nu_U=3 \nu_L$). We also consider here the pair
($\alpha$, $\beta$), being only equal to ($\theta$, $r$) and (K, $r$), assuming $\nu_{\alpha}$ > $\nu_{\beta}$. Note that
CS.1. would lead to $\nu_{\alpha}$ < $\nu_{\beta}$ and is therefore not considered. The last column gives the
corresponding mass for a Schwarzschild black hole ($a=0$), for which $\nu_{\theta}=\nu_\textrm{K}$.

\begin{table}
\centering
\caption{Direct and simple combinational  (beat) resonances matching the observed 3:1 frequency ratio. Last column gives the corresponding mass for a Schwarzschild black hole.}
\label{tab:reson}
\begin{tabular}[t]{| l l l  l l |}
\hline
Type & $\nu_{U}$ & $\nu_{L}$ & n $\nu_{\beta}$ =m $\nu_{\alpha}$ & M$_{a=0}$\\
 & & & n : m & x1e7 M$_{\odot}$\\
\hline
D1-2 &$\nu_{\alpha}$ & $\nu_{\beta}$ & 3 : 1  & 15.8\\
CS.2. & $\nu_{\alpha}$ & $\nu_{\alpha}-\nu_{\beta}$ &  3 : 2 & 7.8\\
CS.3. & $\nu_{\alpha}+\nu_{\beta}$ & $\nu_{\beta}$ & 2 : 1 & 18.3\\
CS.4. & $\nu_{\alpha}-\nu_{\beta}$ & $\nu_{\beta}$ & 4 : 1 & 12.8 \\
CS.5. & $\nu_{\beta}$ & $\nu_{\alpha}-\nu_{\beta}$ & 4 : 3 & 4.1\\
CS.6. &  $\nu_{\alpha}+\nu_{\beta}$ & $\nu_{\alpha}-\nu_{\beta}$ & 2 : 1 &18.3\\
\hline
\end{tabular}
\end{table}

Fig.~\ref{fig:mass_spin} shows the mass-spin dependence for the direct and simple combinational (beat) resonances, following the definition of Table~\ref{tab:reson} with  $\nu_{\alpha}=\nu_{\theta}$ (top) and   $\nu_{\alpha}=\nu_{K}$ (bottom).

The 4:3 resonance between the epicyclic frequencies provide the lower mass for the black hole. It is marginally compatible with the black hole mass derived from the SED, assuming the spin is close to $-1$ (retrograde spin).

\section{Conclusions}
\xmmuj\ is now confirmed to be spatially coincident with a Seyfert 2 galaxy for which we have discovered twin peak QPOs. These periodic oscillations have been reported for only few AGNs in the last years, although they were predicted in 2005. This is mainly because periods are expected to be much longer than for galactic black holes and most of them can't be observed within a single observation from any X-ray observatory. Twin peak QPOs are even more rare and are most of the time found in a 3:2 ratio while both frequencies of \xmmuj\ are in a 3:1 ratio. Thanks to the theory of resonances from epicyclic frequencies we managed to provide mass-spin relations, which means that a precise value of the black hole mass will constrain the value of its spin. We provide some hint of its mass via the modelling of its spectral energy distribution, although the uncertainty on the mass is still quite large due to a lack of data in several energy/wavelength bands.

\xmmuj\  has been reported into the literature by \cite{Carpano2008} because of the discovery of a flux drop of a factor of 6.5 within one hour in its first XMM-Newton observation. No such rapid and persistent flux drop or enhancement has been noticed in any subsequent XMM-Newton observation. It is still possible this is linked to a peculiar event like for example the merger of two neutrons stars producing a rapidly spinning magnetar and then collapsing into a black hole. 

More X-ray observations are necessary to understand the source properties better and to confirm the presence or absence of any further sharp flux drop or enhancement.

\section*{Acknowledgements}  
The scientific results reported in this article are based on data obtained with the Swift gamma-ray burst mission, from the Chandra Data Archive and on observations obtained with XMM-Newton, an ESA science mission with instruments and contributions directly funded by ESA Member States and NASA. We also thank Dr. Peter Jonker for triggering Chandra and Swift observations of \xmmuj.

%%%%%%%%%%%%%%%%%%%%%%%%%%%%%%%%%%%%%%%%%%%%%%%%%%


\bibliographystyle{mnras}
 \bibliography{article}
 
% Don't change these lines
\bsp	% typesetting comment
\label{lastpage}
\end{document}